%% file: main.tex
\documentclass[conference,10pt]{IEEEtran}
\usepackage[switch]{lineno}
\usepackage{cite}
\usepackage{fancyhdr}
\usepackage[table,xcdraw]{xcolor}
\usepackage{subcaption}
\usepackage{listings}
\usepackage{courier}
\usepackage{graphicx}
\usepackage{arydshln}
\usepackage{url}
\usepackage{hyperref}
\hypersetup{
 colorlinks=true,
 linkcolor=blue,
 filecolor=blue,
 citecolor = blue,      
 urlcolor=blue,
}
\usepackage{multicol}
\definecolor{codegreen}{rgb}{0,0.6,0}
\definecolor{codegray}{rgb}{0.5,0.5,0.5}
\definecolor{codepurple}{rgb}{0.58,0,0.82}
\definecolor{backcolour}{rgb}{0.95,0.95,0.92}

\lstdefinestyle{mystyle}{
    backgroundcolor=\color{backcolour},   
    commentstyle=\color{magenta},
    keywordstyle=\color{magenta},
    numberstyle=\tiny\color{codegray},
    stringstyle=\color{codepurple},
    basicstyle=\ttfamily\footnotesize,
    breakatwhitespace=false,         
    breaklines=true,                 
    captionpos=b,                    
    keepspaces=true,                 
    numbers=left,                    
    numbersep=5pt,                  
    showspaces=false,                
    showstringspaces=false,
    showtabs=false,                  
    tabsize=2
}
\usepackage{xcolor}
\newcommand{\tcolR}{\textcolor{red}}
\newcommand{\tcolB}{\textcolor{blue}}
\definecolor{navyblue}{rgb}{0.0, 0.0, 0.5}
\newcommand{\tcolNB}{\textcolor{navyblue}}
\usepackage{amssymb,amsfonts}
\usepackage[numbers]{natbib}
\usepackage{amsmath}
\usepackage{graphicx}
\usepackage{booktabs}
\usepackage{textcomp}
\usepackage{amsmath,amssymb,amsfonts}
\usepackage{graphicx}
\usepackage{textcomp}
\usepackage{xcolor}
\usepackage{etoolbox}
\usepackage{algorithm}
\usepackage{algpseudocode}
\usepackage{lipsum}
\usepackage{perpage} 
\MakePerPage{footnote} 

\usepackage{csquotes}

\DeclareMathOperator*{\argmax}{arg\,max}

\newcommand{\setRequirements}{\mathcal{R}}
\newcommand{\setRequirementVectors}{\vec{\setRequirements}}
\newcommand{\setCandidateConflicts}{\mathcal{C}}
\newcommand{\matrixCosineSimilarity}{\Delta}
\newcommand{\setFinalConflicts}{\bar{\setCandidateConflicts}}
\newcommand{\candidateConflict}{c}
\newcommand{\requirement}{r}

\newcommand{\nsimilar}{m}
\newcommand{\setSimilarRequirement}{\mathcal{L}}

\newcommand{\overlapthreshold}{T_o}
\newcommand{\overlapratio}{v}
\newcommand{\cosineSimilarityCutoff}{\delta}

\newcolumntype{C}[1]{>{\centering\arraybackslash}p{#1}}
\newcolumntype{P}[1]{>{\raggedright\arraybackslash}p{#1}}
\def\BibTeX{{\rm B\kern-.05em{\sc i\kern-.025em b}\kern-.08em
    T\kern-.1667em\lower.7ex\hbox{E}\kern-.125emX}}
\newenvironment{RQquestion}{%
  \par%
  \leftskip=1em\rightskip=1em%
  \noindent
  \ignorespaces}
  
\newenvironment{RQanswer}{%
  \par%
  \leftskip=3em\rightskip=2em%
  \noindent
  \ignorespaces}{%
  \par\medskip}

\begin{document}

\title{Supervised Semantic Similarity-based Conflict Detection Algorithm: S3CDA
}

\author{
\IEEEauthorblockN{Garima Malik}
\IEEEauthorblockA{Toronto Metropolitan University\\
Toronto, Ontario\\
garima.malik@torontomu.ca}
\and
\IEEEauthorblockN{Mucahit Cevik}
\IEEEauthorblockA{Toronto Metropolitan University\\
Toronto, Ontario\\
mcevik@torontomu.ca}
\and
\IEEEauthorblockN{Ayşe Başar}
\IEEEauthorblockA{Toronto Metropolitan University\\
Toronto, Ontario\\
ayse.bener@torontomu.ca}
\and
\IEEEauthorblockN{Devang Parikh}
\IEEEauthorblockA{IBM Canada\\
dparikh@us.ibm.com}
}

\maketitle

\begin{abstract}
Identifying conflicting requirements is a key challenge in software requirement engineering, often overlooked in automated solutions. Most existing approaches rely on handcrafted rules or struggle to generalize across different domains. In this paper, we introduce S3CDA, a two-phase algorithm designed to automatically detect conflicts in software requirements. Our method first identifies potentially conflicting requirement pairs using semantic similarity, and then validates them by analyzing overlapping domain-specific entities. We evaluate S3CDA on five diverse real-world datasets and compare it against popular large language models like GPT-4o, Llama-3, Sonnet-3.5 and Gemini-1.5. While LLMs show promise, especially on general datasets, S3CDA consistently performs better in domain-specific settings with higher performance. Our findings suggest that combining Natural Language Processing (NLP) techniques with domain-aware insights offers a practical and effective alternative for conflict detection in requirements.
\end{abstract}


\section{Introduction}
The term \textit{Conflict} in requirement engineering domain lacks a universally accepted definition and is often domain-dependent. Various studies came up with their own set of definitions like conflicts involves interference, interdependency, or inconsistency between requirements. \citet{kim2007managing} define conflicts as interactions or dependencies that cause undesired software behavior. Others have categorized conflicts by requirement type \citep{butt2011requirement}, complexity \citep{moser2011requirements}, and semantic relationships, where inferential, interdependent, or inclusive links between functional requirements may lead to inconsistent system behavior \citep{guo2021automatically}.

In our work, the conflicts are defined based on the premise that if the implementation of $\requirement_i$ and $\requirement_j$ cannot coexist or if the implementation of the first adversely impacts the second, then they are considered conflicts in our dataset. For instance:
\begin{enumerate}\setlength\itemsep{0.3em}
    \item The UAV shall charge to 50 \% in less than 3 hours.
    \item The UAV shall fully charge in less than 3 hours.
\end{enumerate}
It's not feasible to implement these requirements simultaneously as it will lead to inconsistency in the system. Notably, for the purpose of this study, requirements deemed as duplicates or paraphrased versions of each other are also considered conflicts due to their inherent redundancy. 

Existing studies for conflict detection either rely on handcrafted rules or struggle with domain adaptation. Additionally, many of these approaches are evaluated on limited datasets, raising concerns about their generalizability (refer Table~\ref{tab:conflict_lit_1}) . To address these limitations, we propose a supervised two-phase framework called Supervised Semantic Similarity-based Conflict Detection Algorithm (S3CDA). Our method automatically infers conflict in requirements and outputs conflict/non-conflict labels. In Phase I, we represent each requirement as a high-dimensional sentence embedding and identify potential conflicts based on cosine similarity. In Phase II, we refine this candidate set by analyzing entity-level overlaps between requirement pairs, with higher overlap indicating a stronger likelihood of conflict. The research is guided by the following key questions: 
\medskip
\begin{RQquestion}
    \textbf{RQ1: Which text embeddings are most effective for detecting conflicts in software requirements?}
\end{RQquestion}
\medskip
\begin{RQanswer}
    We evaluate the effectiveness of various text embeddings for conflict detection in software requirements. As baselines, we implement Term Frequency–Inverse Document Frequency (TF-IDF) and Universal Sentence Encoder (USE), and compare them with Sentence-BERT (SBERT). To leverage both semantic and frequency-based features, we propose a hybrid embedding that combines SBERT and TF-IDF. Additionally, we assess PUBER~\citep{das2021sentence}, a domain-specific embedding model trained on software requirements in an unsupervised manner. For conflict detection, we compute cosine similarity between embedded requirements and determine a data-driven similarity threshold to identify potential conflicts. All embeddings are evaluated across five different requirements datasets.
\end{RQanswer}

\medskip
\begin{RQquestion}
    \textbf{RQ2: Which NER method is most effective in identifying overlapping entities within software requirements?}
\end{RQquestion}
\medskip
\begin{RQanswer}
   To further validate potential conflicts between requirements, we introduce a Named Entity Recognition (NER)-based approach for extracting key entities. We compare two NER methods: (1) a software-specific NER model trained to identify domain-relevant entities such as actors, actions, objects, and metrics, and (2) a generic NER model that extracts common linguistic elements like nouns and verbs. For each potentially conflicting requirements, we first determine a set of similar requirements and then we compute the overlap between their extracted entities. A high degree of overlap indicates a stronger likelihood of actual conflict. The effectiveness of each NER method is evaluated based on how well this entity overlap correlates with true conflicts in the dataset. 
\end{RQanswer}

\medskip
\begin{RQquestion}
    \textbf{RQ3: How do state-of-the-art LLMs compare to the S3CDA algorithm in terms of conflict detection performance?}
\end{RQquestion}
\medskip
\begin{RQanswer}
    The advent of large language models (LLMs) has significantly impacted various natural language processing tasks, including conflict detection in software requirements. However, performance and practical applicability of these large models remains an open question. This study compares open-source and third-party LLMs like GPT-3.5~\citep{openai2023gpt35}, GPT-4o~\citep{openai2023gpt4}, Sonnet-3.5~\citep{anthropic2024}, Gemini-1.5~\citep{gemini2024}, and Llama-3~\citep{touvron2023llama} in
    zero-shot settings. We evaluate the performance of these LLMs with all the requirement datasets for conflict detection.
\end{RQanswer}

\section{Literature Review}
Conflict detection is one of the most difficult problems in requirement engineering \citep{aldekhail2016software}. Inability in identifying the conflicts in software requirements might lead to uncertainties and cost overrun in software development. Several papers discussed conflict detection in various domains, however, an autonomous, reliable and generalizable approach for detecting conflicting requirements is yet to be achieved. Below, we review the conflict detection strategies for functional and non-functional requirements.

Functional requirements define the specific behaviors, features, or operations that a system must support. They describe what the system should do under various conditions and form the foundation of system specifications. Table~\ref{tab:conflict_lit_1} summarizes recent studies that focus on conflict detection in functional requirements, highlighting the detection methods, domains, datasets, and conflict types addressed. The majority of studies in Table~\ref{tab:conflict_lit_1} apply rule-based, heuristic, or logic-based techniques. These methods often rely on handcrafted rules, predefined patterns, or logical inference to detect inconsistencies or overlaps in requirements. However, most of these approaches are evaluated using limited datasets—typically small case studies or domain-specific collections—posing challenges for replicability and generalizability. The absence of standard benchmarks and consistent definitions of conflict further complicates comparison across studies.

\renewcommand{\arraystretch}{1.3}
\begin{table*}[!ht]
    \centering
    \caption{Conflict detection literature for SRS documents.}
    \label{tab:conflict_lit_1}
    \resizebox{\linewidth}{!}{
    \input{Tables/conflict_lit_1}}   
\end{table*}

Notably, \citet{guo2021automatically} proposed a more comprehensive and systematic approach called FSARC (Finer Semantic Analysis\-based Requirements Conflict Detector). FSARC integrates Stanford CoreNLP~\citep{manning2014stanford} for linguistic preprocessing, including POS tagging and dependency parsing, and transforms each requirement into an eight-element tuple: \texttt{(id, group\_id, event, agent, operation, input, output, restriction)}. This structured representation enables semantic-level comparison between requirements. Rule-based logic is then applied to detect conflicts arising from overlapping or contradictory tuple components. While FSARC shows promise for generalization, its reliance on accurate NLP parsing and a rigid requirement structure can limit robustness in noisy or variably written SRS documents.

Other studies focus on narrower aspects of functional conflict. For example, \citet{aldekhail2022intelligent} combine rule-based methods with Genetic Algorithms (GA) to resolve conflicts in three SRS documents, while \citet{kamalrudin2010managing} detect conflicts in Essential Use Cases using traceability links and dependency analysis. \citet{viana2017identifying} use fuzzy temporal logic to handle resource-based conflicts in IoT systems, and \citet{felty2003feature} apply linear temporal logic in the telecom domain to reason about feature interactions. These diverse approaches illustrate the evolving nature of conflict detection, though most are still limited by the scale and domain specificity of their evaluations.

Non-functional requirements (NFRs) define how a system should perform, focusing on attributes like reliability, scalability, security, and usability. As shown in Table~\ref{tab:conflict_lit_1}, most conflict detection studies for NFRs rely on rule-based or manual methods and are often evaluated on small datasets, limiting generalizability. Some recent work, such as \citet{abeba2021identification}, applies machine learning (SVM and BiLSTM with Word2Vec) to detect quality-related conflicts, though scalability remains a concern. Other studies target specific conflict types-e.g., OAM\&P conflicts \citep{chentouf2014managing} or mutually exclusive attributes \citep{sadana2007analysis}-but often lack integration into broader conflict resolution workflows.

In our study, we aim to develop an automated NLP-based conflict detection algorithm capable of handling variably structured and written software requirements across different domains. Unlike existing approaches that often rely on rigid structures or domain-specific rules, our method is designed to generalize across diverse requirement styles. Table~\ref{tab:conflict_lit_1} outlines the key characteristics of our proposed approach and highlights how it differs from prior work.

\section{Methodology}\label{sec:method}
In this section, we first describe the SRS datasets used in our numerical study. Then, we provide specific details of the building blocks of our conflict detection algorithms and the experimental setup. 
\subsection{Datasets}\label{subsec:data_sets}
We use five SRS datasets spanning domains such as software, healthcare, transportation, and hardware. Three are open-source (OpenCoss, WorldVista, and UAV), while the other (PURE and IBM-UAV) two are extracted from public or proprietary sources. To ensure structural consistency, we retain the original format of each requirement and convert complex requirements such as paragraphs involving multiple actors and actions into simpler requirements. Since the open-source datasets contain limited \textit{known conflicts}, we introduced synthetic conflicts with guidance from system engineering experts at IBM. For the proprietary IBM-UAV dataset, experts provided synthetic conflicts modeled after real conflicts observed in IBM’s requirement management products. (See Table~\ref{tab:datasets} for summary statistics.)
\renewcommand{\arraystretch}{1.3}
\begin{table}[!ht]
    \centering
    \caption{Dataset characteristics. }
    \label{tab:datasets}
    \resizebox{\linewidth}{!}{
    \input{Tables/dataset_stats}}   
\end{table}
\begin{itemize}\setlength\itemsep{0.6em}
    \item \textbf{OpenCoss}: A safety-critical system dataset covering railway, avionics, and automotive domains\footnote[1]{http://www.opencoss-project.eu}. 
    
    \item \textbf{WorldVista}: A healthcare system capturing patient-related processes\footnote[2]{http://coest.org/datasets}.
    
    \item \textbf{UAV}: Developed at the University of Notre Dame \cite{guo2021automatically,cleland2018dronology}, this dataset uses the EARS template \cite{mavin2009easy} for functional requirements of UAV control systems.
    
    \item \textbf{PURE}:  A subset of the PURE dataset \cite{ferrari2017pure}, containing 83 requirements extracted from two public SRS documents (THEMAS and Mashbot).
    \item \textbf{IBM-UAV}: A proprietary dataset from IBM, covering aerospace and automotive domains. Requirements follow IBM’s RQA (Requirement Quality Analysis) format.
\end{itemize}
Table~\ref{tab:data_sample} presents sample conflicts from the WorldVista, IBM-UAV, and PURE datasets. Each dataset follows a consistent structure with columns for requirement ID, requirement text, a `Conflict' label (Yes/No), and a `Conflict-Label' indicating the conflicting requirement pair. In this study, we focus only on one-to-one conflicts, where each requirement is paired with at most one conflicting counterpart.
\renewcommand{\arraystretch}{1.5}
\begin{table*}[!ht]
\centering
    \caption{Examples of conflicting requirements from requirement datasets.}
    \resizebox{0.9\linewidth}{!}{
    \input{Tables/data_sample}
    }
    \label{tab:data_sample}
\end{table*}
Additionally, we manually annotated the datasets to identify the specific types of conflicts our method aims to detect. This analysis revealed four primary conflict types in the annotated requirements: Security, Functional, Metric, and Operator conflicts as seen in Table~\ref{tab:type_conflicts}.
\begin{table}[!ht]
\centering
    \caption{Type of conflicts in each dataset}
    \label{tab:type_conflicts}
    \resizebox{\linewidth}{!}{
    \input{Tables/conflict_stats}}
\end{table}
Below, we define each conflict type and summarize their distribution across datasets. 
\begin{itemize}\setlength\itemsep{0.6em}
    \item \textbf{Security}: Arise when security requirements contradict, such as unrestricted access vs. strict controls. Found in WorldVista (12) and IBM-UAV (2).
    \item \textbf{Functional}: Involve contradictory system behaviors or functionalities. Most common, especially in WorldVista (46) and PURE (34)
    \item \textbf{Metric}: Stem from inconsistencies in numerical values (e.g., time, distance), requiring semantic interpretation
    \item \textbf{Operator}: Result from conflicting logical terms like ``at least'' vs. ``not more than.''
\end{itemize}
Functional conflicts are the most prevalent, scoping the complexity and breadth of our conflict detection approach.

\paragraph*{Overlapping Entity Ratio vs Cosine Similarity}\label{subsubsec:oer_cos}
Cosine similarity and overlapping entity ratio are central to the effectiveness of our proposed approach, S3CDA. To understand their relationship across various conflict types, we compute the average values of cosine similarity and overlapping entity ratio using Equation~\ref{eq:overlapping} for each dataset and conflict category. Table~\ref{tab:avg_cs_or} presents these aggregated values, offering insights into how different types of conflicts are captured. Overall, cosine similarity proves to be a strong indicator for identifying semantic conflicts, particularly in functional and metric-related requirements. However, in cases like operator or security conflicts—where textual structure or entity presence plays a more prominent role—the overlapping entity ratio offers critical complementary insight. For instance, security conflicts in WorldVista and IBM-UAV exhibit high overlap ratios (0.84 and 0.89), reinforcing the need for entity-level understanding beyond semantics. Also, functional conflicts consistently show high values for both cosine similarity and overlap ratio across all datasets, indicating that these two signals align well in detecting functional conflicts. These averages reinforce the strength of combining both metrics: while cosine similarity captures semantic alignment, overlapping entities improve interpretability and help detect syntactic and structural conflicts that embeddings alone might miss.

\begin{table}[!ht]
\centering
\caption{Average cosine similarity and overlapping entity ratio across conflict types and datasets.}
\label{tab:avg_cs_or}
\resizebox{\linewidth}{!}{
\begin{tabular}{l|ccccc}
\toprule
 &  \multicolumn{5}{c}{Avg. Cosine Similarity / Avg. Overlapping Entity Ratio} \\
\cmidrule(lr){2-6}
\textbf{Conflict Type} & \textbf{OpenCoss} & \textbf{WorldVista} & \textbf{UAV} & \textbf{PURE} & \textbf{IBM-UAV} \\
\midrule
Security& -& 0.82 / 0.84 &- & - & 0.92 / 0.89   \\
Functional &0.97 / 0.96 & 0.85 / 0.86 & 0.95 / 0.91 & 0.95 / 0.91 & 0.92 / 0.96\\
Metric &- & 0.78 / 0.73  & 0.94 / 0.83 &- & 0.96 / 0.96 \\
Operator & -& 0.82 / 0.86  &- & 1.00 / 0.97 &- \\
\bottomrule
\end{tabular}
}
\end{table}


\subsection{S3CDA: Supervised Semantic Similarity-based Conflict Detection Algorithm}
\begin{figure*}[!ht]
    \centering
    \includegraphics[width=1.0\textwidth]{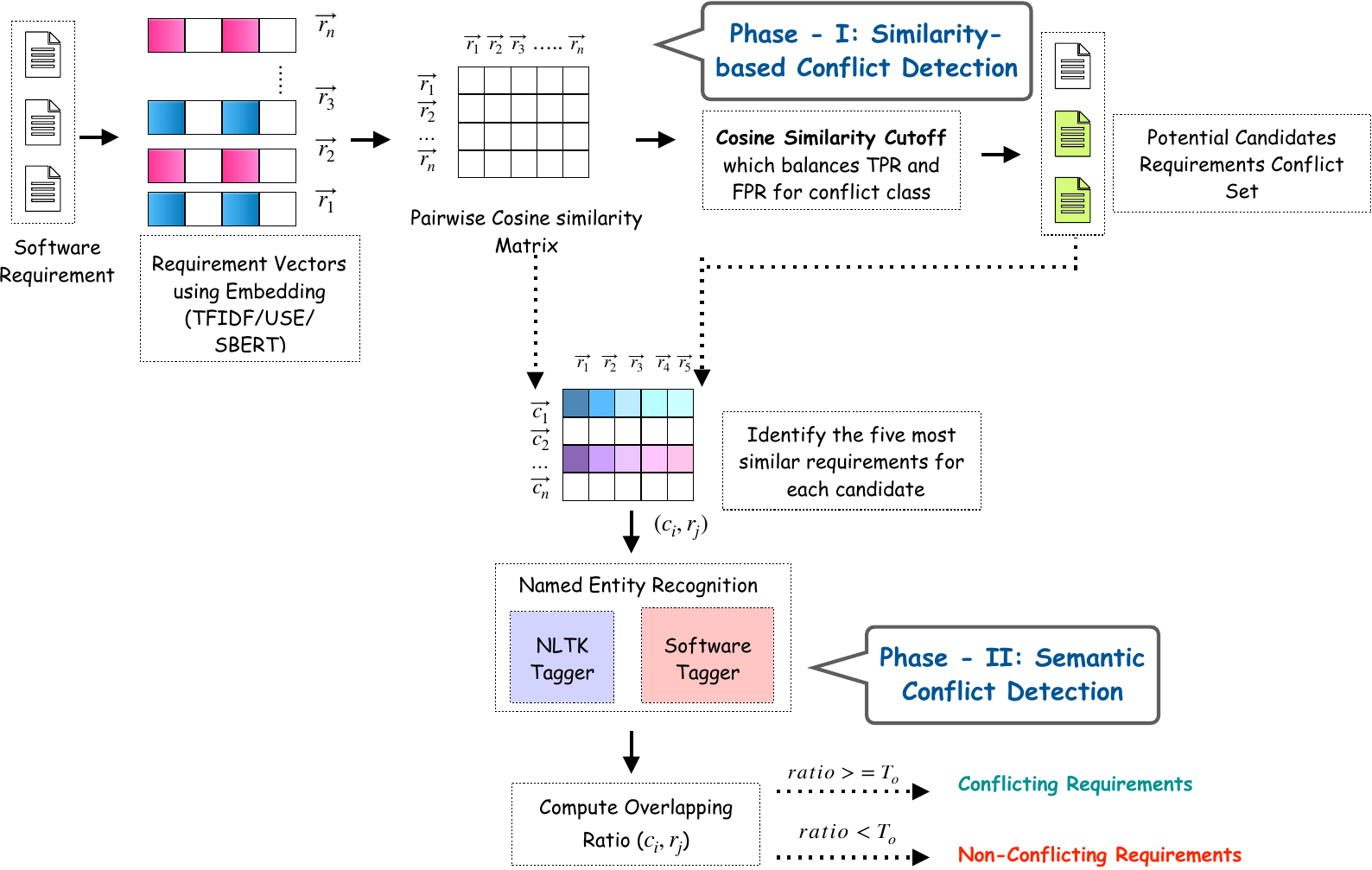}
    \caption{Overview of the S3CDA. \textbf{Phase-I}: identifies potential conflicts using sentence embeddings, cosine similarity, and a learned cutoff. \textbf{Phase-II}: refines these by retrieving similar requirements, extracting entities, computing overlap ratios, and classifying final conflicts.}
    \label{fig:s3cda}
\end{figure*}

This section is structured into two parts as depicted in Figure~\ref{fig:s3cda}. First, we explain the similarity-based conflict detection. Second, we define the semantic-based conflict detection to validate the potential conflicts obtained in Phase I.

\paragraph{Phase I: Similarity-based Conflict Detection}
Algorithm~\ref{alg:pcs} describes the first phase of S3CDA, as illustrated in Figure~\ref{fig:s3cda}. The goal of this phase is to identify candidate conflicts by measuring the similarity between requirements using sentence embeddings and cosine similarity.
We first create the sentence embedding vector for each requirement $\requirement \in \setRequirements$ using \texttt{GenerateEmbedding($\setRequirements$)} procedure. It basically converts the requirements into numerical vector using word/sentence embeddings. Below, we describe the various sentence embedding models employed in the proposed algorithms.
\begin{itemize}\setlength\itemsep{0.6em}
    \item \textbf{TF-IDF}: Captures word importance based on frequency within and across requirements..
    \item \textbf{USE}: Generates 512-dimensional vectors using Universal Sentence Encoder, capturing semantic meaning. 
    \item \textbf{SBERT+TF-IDF}: This method combines SBERT~\citep{reimers2019sentence} and TF-IDF embeddings. SBERT provides context and semantics in the vectors, while TF-IDF prioritizes less frequent words. The process involves concatenating the vectors and employing Uniform Manifold Approximation and Projection (UMAP) for dimensionality reduction to ensure uniform vector size \citep{mcinnes2018umap}.
\end{itemize}
{\footnotesize
\begin{algorithm}[!ht]
\caption{Similarity-based Conflict Detection}\label{alg:pcs}

\textbf{Require:} \\
\hspace*{1em} Requirement set $\setRequirements = \{\requirement_1, \requirement_2, \hdots, \requirement_n\}$ \\
\textbf{Output:} \\
\hspace*{1em} Conflict set $\setCandidateConflicts$

\begin{enumerate}
    \item $\setRequirementVectors \leftarrow \texttt{GenerateEmbeddings}(\setRequirements)$ \\
    \item $\matrixCosineSimilarity_{n \times n} \leftarrow \texttt{ComputeCosineSimilarity}(\setRequirementVectors)$ \\
    \item $\cosineSimilarityCutoff \leftarrow \texttt{FindOptimalCutoff}(\matrixCosineSimilarity, \setRequirements)$ \\
    \begin{equation*}
        \cosineSimilarityCutoff = \argmax_{k \in \left\{ 0.01, \hdots, 1.00 \right\}} \left[ \text{TPR}(k) - (1 - \text{FPR}(k)) \right]
    \end{equation*}
where $\text{TPR}(k)$ and $\text{FPR}(k)$ are computed for each threshold $k$.
    \item For each test set requirement $\requirement_i$:
    \begin{itemize}
        \item Find its closest match $\requirement_j \in \setRequirements, i \neq j$ using:
        \begin{equation*}
            \requirement_j = \argmax_{j} \left\{ \matrixCosineSimilarity[i][j] < 1 \right\}
        \end{equation*}
        \item If the similarity score exceeds the threshold:
        \begin{equation*}
            \matrixCosineSimilarity[i][j] \geq \cosineSimilarityCutoff
        \end{equation*}
        then mark $(\requirement_i, \requirement_j)$ as a potential conflict.
    \end{itemize}

    \item Collect all such pairs into the candidate conflict set:
    \begin{equation*}
        \setCandidateConflicts \leftarrow \{(\requirement_i, \requirement_j) \, | \, \matrixCosineSimilarity[i][j] \geq \cosineSimilarityCutoff \}
    \end{equation*}
\end{enumerate}

\textbf{Return:} $\setCandidateConflicts$
\end{algorithm}
}
Once the requirements are embedded, we compute a pairwise cosine similarity matrix ($\matrixCosineSimilarity$), where each entry represents the similarity between two requirements. To determine the optimal threshold $\cosineSimilarityCutoff$, we evaluate a range of values $k \in \left\{ 0.01,\hdots,1.00\right\}$ on the training set. The threshold that maximizes the difference between true positive rate (TPR) and false positive rate (FPR) is selected: $\left\{\text{TPR}(k)-(1-\text{FPR}(k))\right\}$
Here, TPR measures the proportion of actual conflicts correctly identified at threshold $k$, while FPR captures the proportion of non-conflicting pairs incorrectly labeled as conflicts. This ensures a balanced trade-off between precision and recall. For each test requirement $\requirement_i$, the algorithm identifies its most similar requirement $\requirement_j$ (excluding self-match) using $\matrixCosineSimilarity$. If $\matrixCosineSimilarity[i][j] \geq \cosineSimilarityCutoff$, $\requirement_i$ is labeled as a conflict. All such pairs form the candidate conflict set for the next phase.

\paragraph{Phase II: Semantic-based Conflict Detection}
The role of this phase is to validate the conflicts obtained in Phase I. Algorithm~\ref{alg:scd} describes the process of semantic conflict detection as presented in right panel of Figure~\ref{fig:s3cda}. 
Specifically, any candidate conflict $\candidateConflict \in \setCandidateConflicts$ is semantically compared against top $\nsimilar$ most similar requirements from $\setRequirements$. This semantic comparison is performed based on overlap ratio between the entities present in the requirements. For a given candidate conflict $\candidateConflict \in \setCandidateConflicts$, overlap ratio is calculated as :
\begin{align}\label{eq:overlapping}
    \overlapratio_{\candidateConflict} = \frac{\max_{\requirement \in \setSimilarRequirement_{\candidateConflict}}\big\{\texttt{Overlap}(\candidateConflict, \requirement) \big\}}{\texttt{UniqueEntities}(\candidateConflict)}
\end{align}
where $\setSimilarRequirement_{\candidateConflict}$ represents the set of $\nsimilar$ most similar requirements to candidate conflict $\candidateConflict$.
The function $\texttt{Overlap}(\candidateConflict, \requirement)$ calculates the number of overlapping entities between 
$\candidateConflict$ and $\requirement$, and function $\texttt{UniqueEntities}(\candidateConflict)$ calculates the number of unique entities in candidate conflict (i.e., a requirement text) $\candidateConflict$. $\texttt{UniqueEntities}$ function excludes `O' tag i.e., non-entity tags or any repeating word which is already tagged. Exact matches of entities are considered for overlap computation. Near matches are not included, ensuring semantic precision. The calculated overlap ratio $\overlapratio_{\candidateConflict}$ for $\candidateConflict \in \setCandidateConflicts$ is then compared against a pre-determined overlap threshold value, $\overlapthreshold$, and $\candidateConflict$ is added to final conflict set $\setFinalConflicts$ if $\overlapratio_{\candidateConflict} \geq \overlapthreshold$.

\begin{algorithm}[!ht]
\caption{Semantic Conflict Detection}\label{alg:scd}
\hspace*{\algorithmicindent} \textbf{Require:}  \\
\hspace*{\algorithmicindent} \hspace*{\algorithmicindent} Candidate conflict set: $\setCandidateConflicts = \{\candidateConflict_1,\candidateConflict_2,\hdots,\candidateConflict_t\}$ \\ 
\hspace*{\algorithmicindent} \hspace*{\algorithmicindent} Requirement set: $\setRequirements = \{\requirement_1, \requirement_2, \hdots, \requirement_n\}$\\
\hspace*{\algorithmicindent} \hspace*{\algorithmicindent} \# of similar requirements: $\nsimilar$   \\
\hspace*{\algorithmicindent} \textbf{Output:} Refined conflict set $\setFinalConflicts$ \\
\hspace*{\algorithmicindent} \textbf{Initialization:}  \\
\hspace*{\algorithmicindent} \hspace*{\algorithmicindent} $\setFinalConflicts = \emptyset$ \hfill// \textit{Initialize conflict set to an empty set}\\ 
\hspace*{\algorithmicindent} \hspace*{\algorithmicindent} $ \overlapthreshold = 1$ \hfill// \textit{Set overlapping threshold as 1}\\
\hspace*{\algorithmicindent} \hspace*{\algorithmicindent} $ \nsimilar = 5$ \hfill// \textit{Set number of similar requirements as 5}\\
\hspace*{\algorithmicindent} \textbf{For} $\candidateConflict \in \setCandidateConflicts$ \textbf{do}: \\
\hspace*{\algorithmicindent} \hspace*{\algorithmicindent} $\setSimilarRequirement$ $\leftarrow$ \texttt{GetSimilarRequirements}($\candidateConflict,\setRequirements, \matrixCosineSimilarity, \nsimilar$) \\
\hspace*{\algorithmicindent} \hspace*{\algorithmicindent} $\overlapratio$ $\leftarrow$ \texttt{GetMaxOverlapRatio}($\candidateConflict,\setSimilarRequirement$) \\
\hspace*{\algorithmicindent} \hspace*{\algorithmicindent} \textbf{If} $\overlapratio \geq \overlapthreshold $ : \\
\hspace*{\algorithmicindent} \hspace*{\algorithmicindent} \hspace*{\algorithmicindent} $\setFinalConflicts \leftarrow \setFinalConflicts \cup\left\{ \candidateConflict \right\}  $ 
\end{algorithm}

To extract the entities from the requirements, we employ two NER techniques described below.
\begin{itemize}\setlength\itemsep{0.6em}
    \item \textbf{Part-of-Speech (POS) Tagging}: \cite{rupp2009requirements} suggest that a software requirement should follow the structure as Actor (Noun) + Object + Action (Verb) + Resource. The generic NER method extracts `Noun' and `Verb' tags from the requirements based on this structure and referred as `POS' tagging. We employ POS tagger provided in SpaCy library in Python. 
    
    \item \textbf{Software-specific Named Entity Recognition (S-NER)}: we use a software-specific NER system to extract entities such as actor, action, object, property, metric, and operator. For example, in the requirement ``\textit{The UAV shall charge to 75\% in less than 3 hours},'' \textit{UAV} is the actor and \textit{charge} is the action. We leverage transformer-based models trained on software texts to perform this extraction effectively \citep{malik2022software}.
\end{itemize}
An important insight for practitioners is that both phases of the proposed approach can be applied independently, depending on data availability. Phase I involves a data-driven optimization step that requires labeled training data to learn the optimal similarity threshold ($\cosineSimilarityCutoff$). However, in the absence of gold-standard labels, Phase II can still be directly applied by computing overlapping entity ratios to identify potential conflicts. While we recommend using both phases iteratively for improved precision and recall, Phase II alone can offer reasonable performance in scenarios where labeled data is unavailable.
\subsection{Large Language Models}
To compare the performance of our proposed algorithm with state-of-the-art LLMs, we use zero-shot learning to get the inference from LLM models. The following prompt structure is used to perform the conflict detection using LLMs for all the requirement datasets. This prompt was optimized using the Promptim library\footnote[3]{https://blog.langchain.com/promptim/}, a prompt engineering tool designed to enhance LLM performance through systematic prompt tuning.
\begingroup
\renewcommand{\thelstlisting}{}
\begin{lstlisting}[language=Python]
System_prompt = """
    You are a requirement engineering specialist.
"""
User_prompt = """
    Label the following requirement as "Yes" if it conflicts with any other requirement in the set, or "No" if it does not conflict with any other requirement. Don't include any explanation.
    Requirement {req_num}: {req_text}
    Label: Yes/No
"""
\end{lstlisting}
\endgroup
Below, we list the LLMs used in our analysis. 
\begin{itemize}
    \item \textbf{GPT-3.5}~\citep{openai2023gpt35}: GPT-3.5 is an advanced language model by OpenAI, known for its strong performance in zero-shot and few-shot learning. It excels in generating coherent text and handling a wide range of natural language tasks.
    \item \textbf{GPT-4o}~\citep{openai2023gpt4}: GPT-4o is the next generation from OpenAI, offering improved reasoning and coherence. It is trained on larger datasets and outperforms GPT-3.5 in complex tasks.
    \item \textbf{Llama-3}~\citep{touvron2023llama}: Llama-3 by Meta AI is designed for efficiency and scalability. It leverages advanced training techniques to achieve strong performance with a focus on adaptability and learning from limited data.
    \item \textbf{Gemini-1.5}~\citep{gemini2024}: Gemini-1.5 is a state-of-the-art model known for its high reasoning capabilities and context understanding. It generates high-quality text and performs robustly across various natural language tasks.
    \item \textbf{Sonnet-3.5}~\citep{anthropic2024}: Sonnet-3.5 by Anthropic focuses on ethical AI and human-AI collaboration. It emphasizes safety and transparency, making it suitable for applications requiring reliable and ethical text generation.
\end{itemize}
\subsection{Experimental Setup}
To evaluate our proposed approaches, we perform 3-fold cross-validation on each requirement dataset. Given the limited size and the distribution of conflicting requirements, each dataset is split into three folds while ensuring that every conflict pair appears within the same fold. This preserves the integrity of conflict relationships during training and testing. Table~\ref{tab:hyper} summarizes the evaluation metrics, model configurations, and hyperparameters used in our experiments.
\renewcommand{\arraystretch}{1.5}
\begin{table}[!ht]
\centering
\caption{Experimental Setup Summary}
\label{tab:hyper}
\resizebox{\linewidth}{!}{
\begin{tabular}{p{0.15\textwidth} p{0.30\textwidth}}
\toprule
\textbf{Aspect} & \textbf{Details} \\
\midrule
\textbf{Cross-Validation} & 3-fold cross-validation (stratified by conflict label) \\
\textbf{Evaluation Metrics} & Macro-averaged Precision, Recall and F1-score\\
\textbf{Embedding Models} & TF-IDF, USE, SBERT(\href{https://huggingface.co/sentence-transformers/all-MiniLM-L12-v2}{all-MiniLM-L12-V2}), SBERT+TF-IDF (concatenated), PUBER~\citep{das2021sentence} \\
\textbf{NER Models} & Software-specific Transformer-based NER~\citep{malik2022software}; Generic \href{https://spacy.io/}{SpaCy} NER \\
\textbf{Hyperparameters} & Language Models: temp=0, output\_tokens=100 \\
\bottomrule
\end{tabular}
}
\end{table}
\section{Results}\label{sec:results}
\begin{figure*}[!ht]
    \centering
    \subfloat[$T_o = 1$\label{fig:label_1}]{\includegraphics[width=0.5\textwidth]{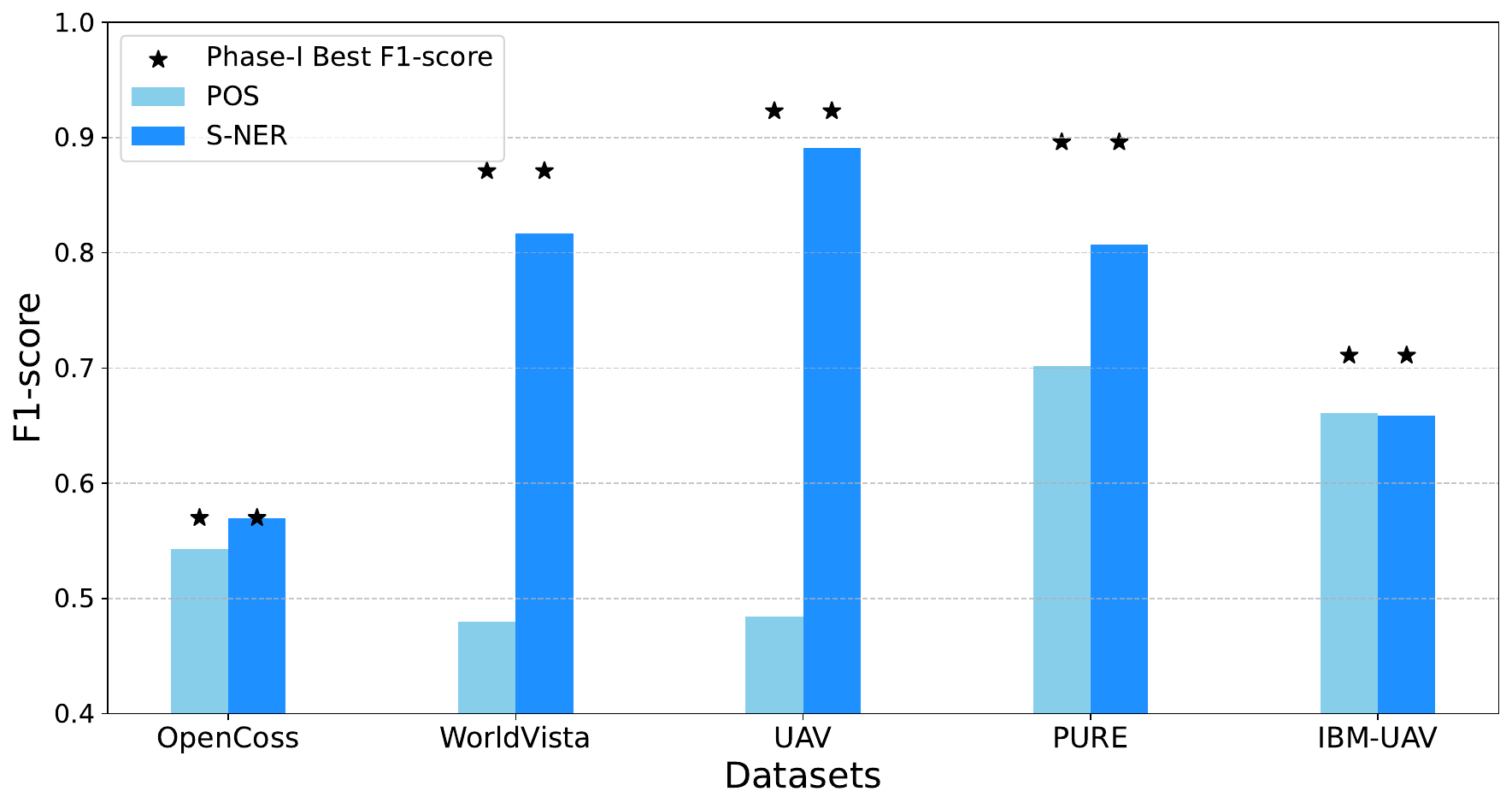}}
    \subfloat[$T_o = 0.75$\label{fig:label_2}]{\includegraphics[width=0.5\textwidth]{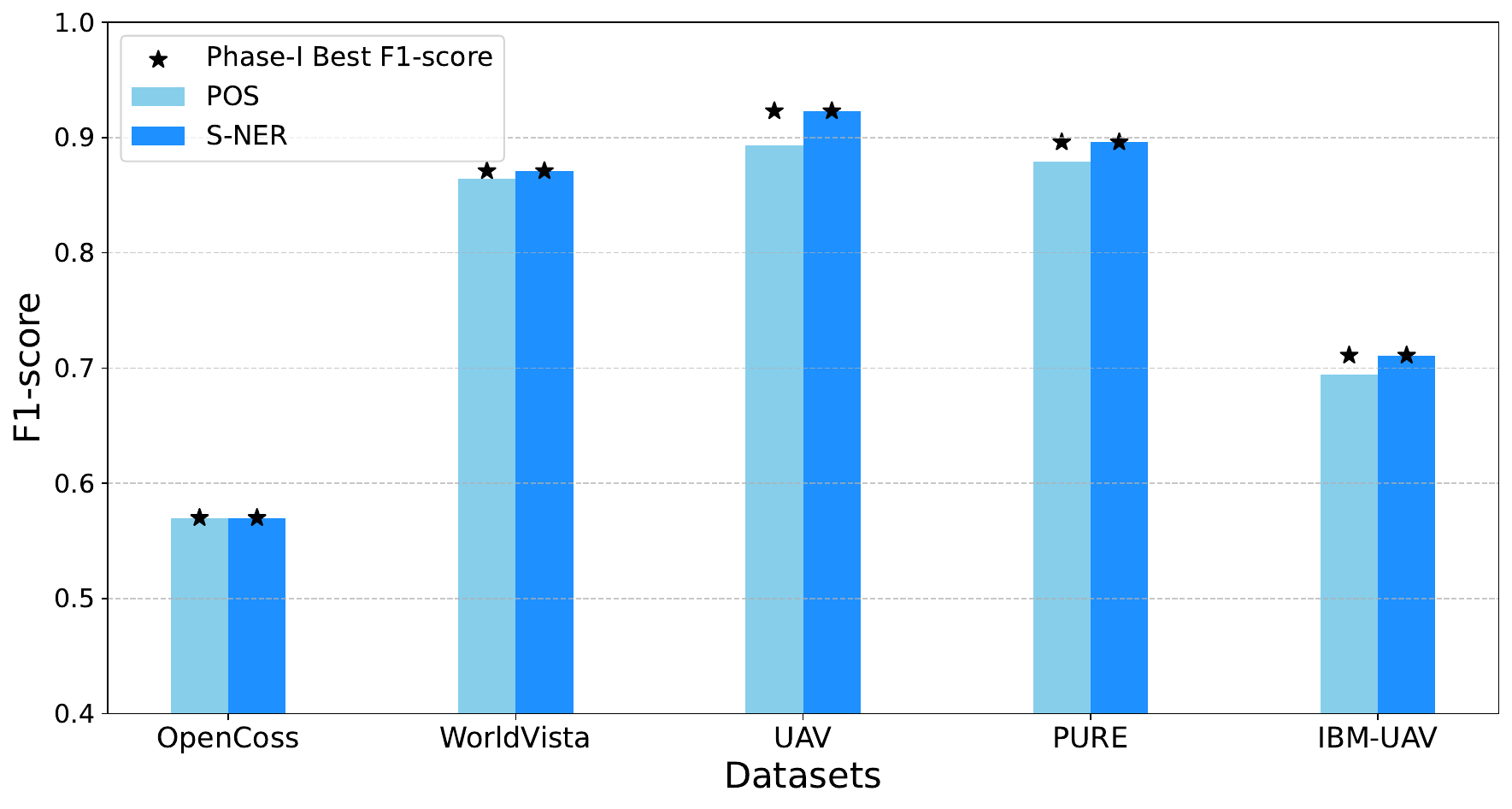}}
    \caption{Macro averaged F1-scores with S-NER and POS entity extraction techniques compared with Phase-I best F1-score for thresholds $T_o = \{1.0,0.75\}$}
    \label{fig:phase_2}
\end{figure*}

In this section, we assess the proposed approach S3CDA for five different requirement datasets along with the comparison with LLMs. 
\medskip
\begin{RQquestion}
    \textbf{RQ1: Which text embeddings are most effective for detecting conflicts in software requirements?}
\end{RQquestion}
Table~\ref{tab:step_1} presents a comparison of various sentence embeddings in Phase I of the S3CDA approach. We observe that SBERT consistently provides the highest F1-score for three out of five datasets. TF-IDF outperforms all other embeddings on OpenCoss (F1 = 0.570), which is atypical. This result suggests that in datasets with high term overlap and less semantic variation, frequency-based models like TF-IDF are more effective than semantic embeddings. USE achieves the best performance on UAV (F1 = 0.923), indicating that deep averaging sentence encoders like USE can be particularly effective in domains with clear and structured requirement templates. WorldVista, which has general healthcare requirements in plain English, shows consistently high performance across all embeddings, with only minor variations in F1 (ranging from 0.821 to 0.871). This indicates that when requirement language is simple and consistent, all embeddings perform well, and choice of embedding may be less critical. SBERT+TF-IDF is competitive across several datasets but never the top performer. It strikes a balance between contextual semantics (via SBERT) and term importance (via TF-IDF), making it a good secondary choice when neither TF-IDF nor SBERT clearly dominates.
\renewcommand{\arraystretch}{1.3}
\begin{table*}[!ht]
\centering
    \caption{Results for various sentence embedding with requirement datasets are reported using macro-averaged metrics - Precision (P), Recall (R), and F1-score (F1) expressed as ``mean ± std" across three folds. The evaluation also includes the number of Potential and Correctly Detected conflicts in comparison to known conflicts for each dataset. Best embedding is highlighted in blue color and yellow highlighting indicates noteworthy results. Here, $E_1$: TF-IDF, $E_2$: USE, $E_3$: SBERT, $E_4$: SBERT+TF-IDF. }
    \label{tab:step_1}
    \resizebox{0.92\linewidth}{!}{
    \input{Tables/phase_1_res}
}
\end{table*}
\medskip
\begin{RQquestion}
    \textbf{RQ2: Which NER method is most effective in identifying overlapping entities within software requirements?}
\end{RQquestion}
Table~\ref{tab:step_2} summarizes the results of phase-II for S3CDA algorithm. S-NER consistently outperforms POS tagging in terms of balanced performance. As expected, a strict threshold ($T_o = 1.0$) results in lower Recall, especially for POS. A flexible threshold ($T_o = 0.75$) mitigates this effect, and in some cases (e.g., UAV and IBM-UAV), allows S-NER to match or outperform Phase I, confirming its value in low-resource or unlabeled settings. The S-NER method is more effective overall in identifying overlapping entities in software requirements. It consistently provides more balanced and reliable performance across datasets and thresholds, particularly in retaining Recall without compromising Precision excessively. While POS tagging may offer gains in isolated cases for Precision, it struggles to generalize and often underperforms in terms of overall conflict coverage. Hence, S-NER is the preferred method for entity extraction in conflict detection pipelines, especially in settings requiring generalizability, robustness to thresholding, and consistent performance across varied requirement styles. 
\renewcommand{\arraystretch}{1.5}
\begin{table*}[!ht]
\centering
\caption{Comparison of POS tagging and S-NER entity extraction techniques with $T_o = \{1.0,0.75\}$. Performance changes are highlighted in blue and red, indicating increments and decrements. The baseline for comparison is established using the best embedding performance as identified in Table~\ref{tab:step_1}.}
\label{tab:step_2}
\resizebox{0.85\linewidth}{!}{
\input{Tables/phase_2_res}

}
\end{table*}
Figure~\ref{fig:phase_2} demonstrates that S-NER consistently outperforms POS, with notable improvements in F1-scores. The strict threshold of $T_o = 1$ proves to be too rigid, often leading to a drop in overall F1-score compared to Phase I. In contrast, using a more flexible threshold ($T_o = 0.75$) enables the S-NER method to match or even surpass Phase I performance, as it effectively captures conflicts with an entity overlap of 75\% or higher.
\begin{RQquestion}
    \textbf{RQ3: How do state-of-the-art LLMs compare to the S3CDA algorithm in terms of conflict detection performance?}
\end{RQquestion}
These results are obtained in a zero-shot setting, where each LLM is directly prompted to label pairs of requirements as conflict or non-conflict, without any task-specific fine-tuning. To ensure consistency with the S3CDA experimental setup, the evaluation is performed across three folds of each dataset. The results in the Table~\ref{tab:compare_llm} demonstrate that S3CDA outperforms or matches LLMs on most datasets. GPT-4o shows the best F1-score among LLMs across multiple datasets (OpenCoss, WorldVista), This is expected due to the multi-model optimization and robust zero-shot learning capabilities for generic requirement datasets. As expected, GPT-3.5 provides average performance as it's not optimized for tasks requiring fine-grained understanding of structured technical language. 
LLaMA-3 (base model ~8B) is open-weight and powerful for fine-tuning, but in this zero-shot context, it lacks task-specific optimization. Sonnet-3.5 and Gemini-1.5, Both models display mid-tier performance—better than GPT-3.5 and LLaMA-3, but lower than GPT-4o and S3CDA. This may be due to their training corpus breadth and instruction-tuning quality, which allows for general reasoning but not enough specialization for software requirements. 

\begin{table*}[!ht]
  \centering
  \caption{Comparison of LLMs and the proposed S3CDA algorithm in terms of conflict detection macro-averaged F1-score across five software requirements datasets. The best-performing method for each dataset is highlighted. Results are reported as ``mean ± std'' over three folds.}
  \label{tab:compare_llm}
  \resizebox{0.92\linewidth}{!}{
  \input{Tables/llm_res}}
\end{table*}
LLMs like GPT-4o, Gemini-1.5, and Claude Sonnet-3.5 are powerful general-purpose models capable of performing various tasks, including conflict detection in software requirements. While they show promising zero-shot results, their full potential may be unlocked with fine-tuning or few-shot learning. However, practical challenges limit their industrial use. First, cost is a major concern. All LLM experiments in our study required paid inference through APIs from OpenAI, Google, and Anthropic. Since pricing is token-based, applying these models to large requirement datasets can become expensive. Second, data privacy is a critical issue. Using LLM APIs means sending data to external servers, which is unsuitable for domains with strict confidentiality requirements. In contrast, our proposed method is cost-efficient, locally deployable, and privacy-preserving. Once trained, it incurs no ongoing costs and can be securely hosted on enterprise infrastructure—making it better suited for long-term, real-world deployment.
\section{Threats to Validity}\label{sec:threats}
Several potential threats to the validity of our conflict detection approach warrant consideration. Since our method relies on cosine similarity, it can struggle with requirements that differ structurally or use varied vocabulary. Below, we highlight specific challenges across conflict types: 
\begin{itemize}
    \item \textbf{Metric and Operator Conflicts}: These involve comparing numerical values and logical phrases (e.g., ``at least," ``not more than"). The model can detect explicit patterns but may fail with unit conversions, implicit logic, or complex conditionals.
    \item \textbf{Security Conflicts}: These require domain-specific context, such as understanding access policies. Embeddings may identify surface-level similarities but often miss deeper policy conflicts or interactions between multiple constraints.
\end{itemize}
The use of manual labeling and synthetic conflicts introduces subjectivity and limits generalizability. Manual annotations, especially for complex conflict types (e.g., metric or security), may vary across annotators. Synthetic conflicts help simulate edge cases but might not fully reflect real-world patterns, especially in specialized domains. Finally, the effectiveness of our approach depends on the accuracy of the software-specific NER model. Errors in entity extraction (e.g., mislabeling actors or actions) directly affect overlap calculations, leading to false positives or negatives. Improving NER performance through fine-tuning on diverse and domain-rich datasets is crucial to mitigate this risk.

\section{Conclusion}\label{sec:conc}
This study addresses the challenge of conflict detection in software requirements by introducing S3CDA, a two-phase approach: Phase-I identifies potential conflicts using sentence embeddings, and Phase-II validates them through entity overlap via software-specific NER. Our evaluation shows that SBERT performs best on semantically rich datasets, while TF-IDF excels in cases with high lexical overlap. The effectiveness of embeddings varies with requirement language—general wording benefits all models, while domain-specific phrasing demands tailored embeddings. In Phase-II, software-specific NER consistently outperforms generic methods for extracting relevant entities. While GPT-4o shows strong performance on general datasets, S3CDA proves more effective for domain-specific conflict detection due to its targeted design.
For future work, we aim to expand to more diverse datasets, explore LLM-based embeddings, enhance the NER module with advanced models, and fine-tune LLMs on requirement texts to assess their generalizability.





\small
\bibliographystyle{IEEEtranN} 
\bibliography{IEEEabrv,Ref}
\end{document}

%% file: Tables/conflict_lit_1.tex
\begin{tabular}{P{0.10\textwidth}P{0.15\textwidth} P{0.20\textwidth} P{0.10\textwidth} P{0.15\textwidth} P{0.15\textwidth}}
\toprule
& {\bf Study}  & {\bf Conflict Detection Method} & {\bf Domain} & {\bf Dataset} & {\bf Type of Conflicts} \\
\midrule
Functional Requirements  & \citet{aldekhail2022intelligent} & Rule-based and Genetic Algorithms (GA) & Software & 3 SRS documents & Functional conflicts   \\
& \citet{guo2021automatically} & Rule-based and semantics & Varied & 3 SRS (validation), 2 SRS (testing) & Functional conflicts \\
& \citet{viana2017identifying} & Fuzzy branching temporal logic & Internet of Things (IOT) & System of Systems (SOS) Req. & Conflicts in resource-based req. \\
& \citet{butt2011requirement}& Automated & Software & 25 Req. & Conflicts in Mandatory, essential and optional req.\\
& \citet{kamalrudin2010managing}  & Rule-based \& Tracing dependencies &Software & Voter registration system & Conflicts in Essential Use Cases (EUC)  \\
\cmidrule{2-6}
Non-Functional Requirements & \citet{shah2021detecting} & Rule-based and Clustering & Software & 15 ATM Req. and 5 SRS documents & Intra-conflicts  \\
& \citet{abeba2021identification}  & SVM and BiLSTM model with Word2Vec embeddings & Software & 200 Req. & Software Product Quality \\
& \citet{chentouf2014managing} & Rule-based and Manual & Telecom & 14 Req. & Req. Conflicts in OAM\&P  \\
& \citet{mairiza2013conflict} & Quantitative analysis & Chemical & Casestudy with 2 req. & \\
& \citet{egyed2004identifying} & Rule-based and Manual & Software & 12 Req. Video on Demand & Req. Traceability \\
\cmidrule{2-6}
\rowcolor{yellow!15}& S3CDA & Automated & Varied & 5 Requirement datasets &  Security, Functional, Metric and Operator\\

\bottomrule
\end{tabular}

%% file: Tables/dataset_stats.tex
\begin{tabular}{p{0.10\textwidth} p{0.10\textwidth} p{0.07\textwidth}p{0.07\textwidth}p{0.07\textwidth}}
\toprule
Dataset & Domain & Non-conflicts & Known Conflicts & Synthetic Conflicts\\
\midrule
OpenCoss   &  Transportation & 97     & 4   & 16   \\
WorldVista &  Medical      &   78     &  10  & 60   \\
UAV         &  Aerospace      &  80    & 10     & 26    \\
\midrule
PURE        &  Thermodynamics &   27  &  0    & 40  \\
IBM-UAV     &  Hardware     &   75  &   0     &  28  \\ 
\bottomrule
\end{tabular}

%% file: Tables/data_sample.tex
    \begin{tabular}{P{0.1\textwidth} P{0.04\textwidth} P{0.53\textwidth} P{0.08\textwidth} P{0.1\textwidth}}
\toprule
Dataset &Req. Id & Requirement text & Conflict & Conflict-Label  \\ 
\midrule
WorldVista & 1. &The system's pilot program shall use a smart card to digitally sign medication orders. & Yes & Yes (2)\\
           & 2. &The system's pilot program shall require a handwritten signature for medication orders. & Yes & Yes (1) \\
           & 3. &The system shall sort notifications based on column heading: Patient name (alphabetical or reverse alphabetical). & No & No \\
\midrule
IBM-UAV & 4.& The UAV shall charge to 50 \% in less than 3 hours. & Yes  & Yes (5)\\
        & 5.& The UAV shall fully charge in less than 3 hours. & Yes & Yes (4)\\
        & 6.& Remote surveillance shall include video streaming for manual navigation of the surveillance platform. & No & No \\
\midrule
UAV & 7. &The \_InternalSimulator\_ shall approximate the behavior of a UAV. & No & No \\
    & 8. &The \_VehicleCore\_ shall support virtual UAVs. & Yes & Yes (9)\\
    & 9. &The \_VehicleCore\_ shall support up to three virtual UAVs. & Yes & Yes (8)  \\
\bottomrule
\end{tabular}

%% file: Tables/conflict_stats.tex
\begin{tabular}{lrrrr}
\toprule
Dataset & \# Security & \# Functional & \# Metric & \# Operator \\
\midrule
OpenCoss   & -   & 20     & -  & -   \\
WorldVista & 12    & 46     & 6   & 6 \\
UAV         & -     & 30     & 6   & -   \\
\midrule
PURE        & -  & 34    & -    & 6  \\
IBM-UAV     & 2  &  14   &   12  & -    \\ 
\bottomrule
\end{tabular}

%% file: Tables/phase_1_res.tex
 \begin{tabular}{ll|cccccc}
\toprule
\textbf{Dataset} & \textbf{Embeddings} & \textbf{Cosine} & \textbf{P} & \textbf{R} & \textbf{F1} & \textbf{Potential} & \textbf{Correctly}\\
& & \textbf{Cutoff} & & & & \textbf{Conflicts} & \textbf{Detected} \\
\midrule
OpenCoss & PUBER~\citep{das2021sentence} & - & 0.515 ± 0.13 & 0.511 ± 0.24 & 0.483 ± 0.17 & - & - \\
& \tcolB{$\mathbf{E_1}$} & 0.70 & 0.433 ± 0.11 & 0.847 ± 0.13 & \cellcolor{yellow!15}\textbf{0.570 ± 0.13}  & 41 / 20 & 17 / 20 \\
& $E_2$ & 0.93 & 0.366 ± 0.05 & 0.652 ± 0.27  & 0.461 ± 0.11  & 34 / 20 & 13 / 20  \\
& $E_3$ &  0.95 & 0.354 ± 0.04 & 0.805 ± 0.14 & 0.484 ± 0.02 & 46 / 20 & 16 / 20 \\
& $E_4$ & 0.82 & 0.373 ± 0.16 & 0.722 ± 0.20 & 0.487 ± 0.18  & 41 / 20 & 14 / 20 \\
\midrule

WorldVista & PUBER~\citep{das2021sentence} & - & 0.861 ± 0.09 & 0.838 ± 0.07 & 0.843 ± 0.08  & - & - \\
& $E_1$ & 0.20 & 0.812 ± 0.09 & 0.853 ± 0.11 & 0.821 ± 0.03 & 76 / 70 & 60 / 70 \\
& $E_2$ & 0.66 & 0.897 ± 0.07 & 0.839 ± 0.11 & 0.857 ± 0.03 & 67 / 70 & 59 / 70 \\
& $E_3$ & 0.76 & 0.856 ± 0.11 & 0.875 ± 0.10 & 0.855 ± 0.05 & 73 / 70 & 61 / 70 \\ 
& \tcolB{$\mathbf{E_4}$} & 0.51 & 0.898 ± 0.08 & 0.856 ± 0.04 & \cellcolor{yellow!15}\textbf{0.871 ± 0.02} & 68 / 70 & 60 / 70\\
\midrule

UAV  & PUBER~\citep{das2021sentence} & - & 0.831 ± 0.09 & 0.863 ± 0.08 & 0.840 ± 0.09 & - & - \\
& $E_1$ & 0.36 & 0.852 ± 0.05 & 0.944 ± 0.07 & 0.894 ± 0.05 & 40 / 36 & 34 / 36\\
& \tcolB{$\mathbf{E_2}$} & 0.77 & 0.860 ± 0.05 & 1.000 ± 0.00 & \cellcolor{yellow!15}\textbf{0.923 ± 0.02} & 42 / 36 & 36 / 36\\
& $E_3$ & 0.88 & 0.918 ± 0.06 & 0.833 ± 0.13 & 0.864 ± 0.06 & 31 / 36 & 30 / 36\\
& $E_4$ & 0.61 & 0.893 ± 0.04 & 0.944 ± 0.04 & 0.917 ± 0.05 & 38 / 36 & 34 / 36 \\
\midrule

PURE & PUBER~\citep{das2021sentence} & - & 0.736 ± 0.06 & 0.743 ± 0.05 & 0.731 ± 0.06 & - & - \\
& $E_1$ & 0.81 & 0.910 ± 0.06 & 0.785 ± 0.21 & 0.819 ± 0.11 & 36 / 40 & 32 / 40 \\
& $E_2$ & 0.81 & 0.893 ± 0.07 & 0.785 ± 0.21 & 0.809 ± 0.10 & 37 / 40 & 32 / 40 \\
& \tcolB{$\mathbf{E_3}$} & 0.88 & 0.902 ± 0.07 & 0.896 ± 0.07 & \cellcolor{yellow!15}\textbf{0.896 ± 0.04} & 40 / 40 & 36 / 40\\
& $E_4$ & 0.71 & 0.958 ± 0.05 & 0.785 ± 0.21 & 0.841 ± 0.12 & 34 / 40 & 32 / 40 \\
\midrule

IBM-UAV & PUBER~\citep{das2021sentence} & - & 0.746 ± 0.01 & 0.715 ± 0.16 & 0.730 ± 0.14 & - & - \\  
& $E_1$ & 0.58 & 0.708 ± 0.21 & 0.550 ± 0.32 & 0.557 ± 0.17 & 24 / 28 & 16 / 28  \\
& $E_2$ & 0.83 & 0.704 ± 0.16 & 0.716 ± 0.33 & 0.688 ± 0.25 & 28 / 28 & 21 / 28 \\
& \tcolB{$\mathbf{E_3}$} & 0.89 & 0.871 ± 0.11 & 0.716 ± 0.33 & \cellcolor{yellow!15}\textbf{0.711 ± 0.22} & 26 / 28 & 21 / 28 \\
& $E_4$ & 0.73 & 0.529 ± 0.38 & 0.566 ± 0.41 & 0.537 ± 0.38 & 22 / 28 & 17 / 28\\
\bottomrule
\end{tabular}

%% file: Tables/phase_2_res.tex
\begin{tabular}{ll|cc|cc}
\toprule
& & \multicolumn{2}{c}{$T_o = 1$} & \multicolumn{2}{c}{$T_o = 0.75$} \\
\cmidrule(lr){3-4} \cmidrule(lr){5-6}
\textbf{Dataset} & \textbf{Method} & \textbf{P} & \textbf{R} & \textbf{P} & \textbf{R} \\
\midrule
OpenCoss & POS & 0.419 ± 0.10 {\scriptsize\tcolR{$\downarrow$0.014}} & 0.791 ± 0.09 {\scriptsize\tcolR{$\downarrow$0.056}} & 0.433 ± 0.12 & 0.847 ± 0.14 \\
         & S-NER & 0.433 ± 0.12 & 0.847 ± 0.14 & 0.433 ± 0.12 & 0.847 ± 0.14 \\
\midrule
WorldVista & POS & 0.925 ± 0.10 {\scriptsize\tcolNB{$\uparrow$0.027}} & 0.327 ± 0.06 {\scriptsize\tcolR{$\downarrow$0.548}} & 0.922 ± 0.08 {\scriptsize\tcolNB{$\uparrow$0.024}} & 0.825 ± 0.08 {\scriptsize\tcolR{$\downarrow$0.050}} \\
           & S-NER & 0.904 ± 0.10 {\scriptsize\tcolR{$\uparrow$0.006}} & 0.756 ± 0.06 {\scriptsize\tcolR{$\downarrow$0.119}} & 0.898 ± 0.09 & 0.856 ± 0.04 \\
\midrule
UAV & POS & 0.925 ± 0.10 {\scriptsize\tcolNB{$\uparrow$0.065}} & 0.361 ± 0.17 {\scriptsize\tcolR{$\downarrow$0.639}} & 0.855 ± 0.04 {\scriptsize\tcolR{$\downarrow$0.005}} & 0.944 ± 0.08 {\scriptsize\tcolR{$\downarrow$0.056}} \\
    & S-NER & 0.867 ± 0.04 {\scriptsize\tcolNB{$\uparrow$0.007}} & 0.916 ± 0.07 {\scriptsize\tcolR{$\downarrow$0.084}} & 0.860 ± 0.05 & 1.000 ± 0.00 \\
\midrule
PURE & POS & 0.850 ± 0.11 {\scriptsize\tcolR{$\downarrow$0.052}} & 0.603 ± 0.12 {\scriptsize\tcolR{$\downarrow$0.293}} & 0.897 ± 0.08 {\scriptsize\tcolR{$\uparrow$0.001}} & 0.869 ± 0.10 {\scriptsize\tcolR{$\downarrow$0.027}} \\
     & S-NER & 0.891 ± 0.08 {\scriptsize\tcolR{$\downarrow$0.011}} & 0.750 ± 0.09 {\scriptsize\tcolR{$\downarrow$0.146}} & 0.902 ± 0.07 {\scriptsize\tcolNB{$\uparrow$0.006}} & 0.896 ± 0.07 \\
\midrule
IBM-UAV & POS & 0.909 ± 0.13 {\scriptsize\tcolNB{$\uparrow$0.038}} & 0.583 ± 0.24 {\scriptsize\tcolR{$\downarrow$0.133}} & 0.864 ± 0.13 {\scriptsize\tcolR{$\downarrow$0.007}} & 0.683 ± 0.31 {\scriptsize\tcolR{$\downarrow$0.033}} \\
        & S-NER & 0.914 ± 0.07 {\scriptsize\tcolNB{$\uparrow$0.043}} & 0.583 ± 0.31 {\scriptsize\tcolR{$\downarrow$0.133}} & 0.871 ± 0.12 & 0.716 ± 0.33 \\
\bottomrule
\end{tabular}

%% file: Tables/llm_res.tex
\begin{tabular}{lrrrrrr}
\midrule
Dataset & \href{https://platform.openai.com/docs/models/gpt-3.5-turbo}{GPT-3.5}~\citep{openai2023gpt35}  & \href{https://openai.com/index/gpt-4o-system-card/}{GPT-4o}~\citep{openai2023gpt4}  & \href{https://huggingface.co/meta-llama/Meta-Llama-3-8B}{Llama-3-7B}~\citep{touvron2023llama} & \href{https://ai.google.dev/gemini-api/docs/models#gemini-1.5-pro}{Gemini-1.5-pro}~\citep{gemini2024} & \href{https://www.anthropic.com/news/claude-3-5-sonnet}{Sonnet-3.5}~\citep{anthropic2024} & S3CDA \\ 
\midrule
OpenCoss   & 0.539 ± 0.15  & \cellcolor{yellow!15}\textbf{0.670 ± 0.14}   & 0.550 ± 0.14 & 0.620 ± 0.13 & 0.590 ± 0.12 & 0.570 ± 0.13 \\ 
WorldVista  & 0.367 ± 0.04  & \cellcolor{yellow!15}\textbf{0.936 ± 0.10}  & 0.400 ± 0.05 & 0.700 ± 0.11 & 0.650 ± 0.10 & 0.871 ± 0.02 \\ 
UAV   & 0.407 ± 0.00  & 0.462 ± 0.07  & 0.430 ± 0.08 & 0.520 ± 0.07 & 0.550 ± 0.06 & \cellcolor{yellow!15}\textbf{0.923 ± 0.02}    \\ 
PURE  & 0.336 ± 0.10  & 0.594 ± 0.05  & 0.380 ± 0.11 & 0.600 ± 0.06 & 0.550 ± 0.05 & \cellcolor{yellow!15}\textbf{0.896 ± 0.04}  \\ 
IBM-UAV  & 0.446 ± 0.04  & 0.634 ± 0.05 & 0.500 ± 0.06 & 0.600 ± 0.05 & 0.650 ± 0.10 & \cellcolor{yellow!15}\textbf{0.711 ± 0.22}  \\ 
\bottomrule
\end{tabular}